\documentclass[preprint]{elsarticle}
\usepackage{graphicx}
\usepackage{amssymb}
\begin{document}
\begin{frontmatter}
\title{Quantum noncomutativity in quantum cosmology}

\author[els]{G. Oliveira-Neto}
%\corref{cor1}\fnref{fn1}}
%\ead{gilneto@fisica.ufjf.br}

\author[gam]{G. A. Monerat}

\author[rvt]{E. V. Corr\^{e}a Silva}

\author[rvt]{C. Neves}

\author[rvt]{L. G. Ferreira Filho}

\address[els]{Departamento de F\'{\i}sica,
Instituto de Ci\^{e}ncias Exatas,
Universidade Federal de Juiz de Fora,
CEP 36036-330 - Juiz de Fora, MG, Brazil.}

\address[gam]{Departamento de Modelagem Computacional,\\
Instituto Polit\'{e}cnico do Rio de Janeiro,\\
Universidade do Estado do Rio de Janeiro,\\
Rua Bonfim, 25 - Vila Am\'{e}lia - Cep 28.625-570,
Nova Friburgo, RJ, Brazil.}

\address[rvt]{Departamento de Matem\'{a}tica, F\'{\i}sica e Computa\c{c}\~{a}o,
Faculdade de Tecnologia,
Universidade do Estado do Rio de Janeiro,
Rodovia Presidente Dutra, Km 298, P\'{o}lo Industrial,
CEP 27537-000, Resende-RJ, Brazil.}


%\fntext[fn1]{phone/fax 55-32-2102-3307(ext. 217)}      

\date{\today}

\begin{abstract}
In the present work, we study the noncommutative version of a quantum
cosmology model. The model has a Friedmann-Robertson-Walker geometry, the 
matter content is a radiative perfect fluid and the spatial sections have 
positive constant curvatures. We work in the Schutz's variational formalism.
We quantize the model and obtain the appropriate Wheeler-DeWitt equation. 
In this model the states are bounded. Therefore, we compute the discrete 
energy spectrum and the corresponding eigenfunctions. The energies depend
on a noncommutative parameter ($\theta$). The solutions to the Wheeler-DeWitt equation
are function of the scale factor ($a$) and a time variable ($\tau$), associated to the fluid.
They also depend on an integer ($n$) and $\theta$. The most general solution ($\Psi(a,\tau)$) to the
Wheeler-DeWitt equation is a sum, in the integer $n$, of the solutions mentioned above.
We observe that, there is no $\Psi(a,\tau)$ satisfying the appropriate boundary conditions.  
Therefore, we conclude that it is not possible to obtain a wavefunction satisfying the appropriate 
boundary conditions for the present model with the considered noncommutativity.
\end{abstract}


\end{frontmatter}

One important arena where noncommutative (NC) ideas may play an important role 
is cosmology. If superstrings is the correct theory to unify all the interactions 
in nature, it must have played the dominant role at very early stages of our 
Universe. At that time, all the canonical variables and corresponding momenta 
describing our Universe should have obeyed a NC algebra. Inspired by these ideas 
some researchers have considered such NC models in quantum cosmology 
\cite{garcia,nelson,barbosa,gil2}. It is also possible that some residual NC contribution 
may have survived in later stages of our Universe. Based on these ideas some 
researchers have proposed some NC models in classical cosmology in order to explain 
some intriguing results observed by WMAP. Such as a running spectral index of the 
scalar fluctuations and an anomalously low quadrupole of CMB angular power spectrum 
\cite{huang,kim,liu,huang1,kim1}. Another relevant application of the NC ideas in 
classical cosmology is the attempt to explain the present accelerated expansion of 
our Universe \cite{pedram,obregon,neves,gil}.

One important arena where noncommutative (NC) ideas may play an important role 
is cosmology. In the early stages of its evolution, the Universe may have had
very different properties than the ones it has today. Among those properties
some physicists believe that the spacetime coordinates were subjected to a 
noncommutative algebra. Inspired by these ideas some researchers have considered 
such NC models in quantum cosmology 
\cite{garcia,nelson,barbosa,gil2}. It is also possible that some residual NC contribution 
may have survived in later stages of our Universe. Based on these ideas some 
researchers have proposed some NC models in classical cosmology in order to explain 
some intriguing results observed by WMAP. Such as a running spectral index of the 
scalar fluctuations and an anomalously low quadrupole of CMB angular power spectrum 
\cite{huang,kim,liu,huang1,kim1}. Another relevant application of the NC ideas in 
classical cosmology is the attempt to explain the present accelerated expansion of 
our Universe \cite{pedram,obregon,neves,gil}.

In the present work, we study the noncommutative version of a quantum
cosmology model. The model has a Friedmann-Robertson-Walker (FRW) geometry, the 
matter content is a radiative perfect fluid and the spatial sections have 
positive constant curvatures. We work in the Schutz's variational formalism 
\cite{schutz,germano1}. The noncommutativity that
we are about to propose is not the typical noncommutativity between usual spatial
coordinates. We are describing a FRW model using the Hamiltonian formalism, therefore
the present model phase space is given by the canonical variables and conjugated
momenta: $\{ a, P_a, \tau, P_\tau \}$. Then, the noncommutativity, at the quantum level, we 
are about to propose will be between these phase space variables. Since these variables 
are functions of the time coordinate $t$, this procedure is a generalization of the typical 
noncommutativity between usual spatial coordinates. The noncommutativity between those
types of phase space variables have already been proposed in the literature. At the quantum
level in Refs. \cite{garcia,nelson,barbosa,gil2}, \cite{obregon} and at the classical level in
Refs. \cite{pedram},\cite{neves,gil}. We quantize the model and obtain the appropriate 
Wheeler-DeWitt equation. In this model the states are bounded. Therefore, we compute the 
discrete energy spectrum and the corresponding eigenfunctions. The energies depend
on a noncommutative parameter ($\theta$). The solutions to the Wheeler-DeWitt equation
are function of the scale factor ($a$) and a time variable ($\tau$), associated to the fluid.
They also depend on an integer ($n$) and $\theta$. The most general solution ($\Psi(a,\tau)$) to the
Wheeler-DeWitt equation is a sum, in the integer $n$, of the solutions mentioned above.
We observe that, there is no $\Psi(a,\tau)$ satisfying the appropriate boundary conditions.  
Therefore, we conclude that it is not possible to obtain a wavefunction satisfying the appropriate 
boundary conditions for the present model with the considered noncommutativity.

The FRW cosmological models are characterized by the
scale factor $a(t)$ and have the following line element,
\begin{equation}  
\label{1}
ds^2 = - N^2(t) dt^2 + a^2(t)\left( \frac{dr^2}{1 - kr^2} + r^2 d\Omega^2
\right)\, ,
\end{equation}
where $d\Omega^2$ is the line element of the two-dimensional sphere with
unitary radius, $N(t)$ is the lapse function and $k$ gives the type of
constant curvature of the spatial sections. Here, we are considering the 
case with positive curvature $k=1$ and we are using the natural
unit system, where $\hbar=c=G=1$. The matter content of the model is
represented by a perfect fluid with four-velocity $U^\mu = \delta^{\mu}_0$
in the comoving coordinate system used. The total energy-momentum tensor 
is given by,
\begin{equation}
T_{\mu,\, \nu} = (\rho+p)U_{\mu}U_{\nu} - p g_{\mu,\, \nu}\, ,  
\label{2}
\end{equation}
where $\rho$ and $p$ are the energy density and pressure of the fluid,
respectively. Here, we assume that $p = \rho/3$, which is the equation of
state for radiation. This choice may be considered as a first approximation
to treat the matter content of the early Universe and it was made as a
matter of simplicity. It is clear that a more complete treatment should
describe the radiation, present in the primordial Universe, in terms of the
electromagnetic field.

From the metric (\ref{1}) and the energy momentum tensor (\ref{2}), one may 
write the total Hamiltonian of the present model ($N {\mathcal{H}}$), where
$N$ is the lapse function and ${\mathcal{H}}$ is the superhamiltonian constraint.
It is given by \cite{germano1},

\begin{equation}
N {\mathcal{H}}= -\frac{p_{a}^2}{12} - 3a^2 + p_{T},  
\label{3}
\end{equation}
where $p_{a}$ and $p_{T}$ are the momenta canonically conjugated to $a$ and 
$T$, the latter being the canonical variable associated to the fluid \cite
{germano1}. Here, we are working in the conformal gauge, where $N = a$. The 
commutative version of the present model was first treated in Ref. \cite{lemos}.

We wish to quantize the model following the Dirac formalism for quantizing
constrained systems \cite{dirac}. First we introduce a wave-function which
is a function of the canonical variables $\hat{a}$ and $\hat{T}$,

\begin{equation}  
\label{4}
\Psi\, =\, \Psi(\hat{a} ,\hat{T} )\, .
\end{equation}
Then, we impose the appropriate commutators between the operators $\hat{a}$
and $\hat{T}$ and their conjugate momenta $\hat{P}_a$ and $\hat{P}_T$.
Working in the Schr\"{o}dinger picture, the operators $\hat{a}$ and $\hat{T}$
are simply multiplication operators, while their conjugate momenta are
represented by the differential operators,
\begin{equation}
p_{a}\rightarrow -i\frac{\partial}{\partial a}\hspace{0.2cm},\hspace{0.2cm} 
\hspace{0.2cm}p_{T}\rightarrow -i\frac{\partial}{\partial T}\hspace{0.2cm}.
\label{5}
\end{equation}

Finally, we demand that the operator corresponding to $N \mathcal{H}$ 
annihilate the wave-function $\Psi$, which leads to the Wheeler-DeWitt 
equation,
\begin{equation}
\bigg(\frac{1}{12}\frac{{\partial}^2}{\partial a^2} - 3a^2
\bigg)\Psi(a,\tau) = -i \, \frac{\partial}{\partial \tau}\Psi(a,\tau),
\label{6}
\end{equation}
where the new variable $\tau= -T$ has been introduced.

The operator $N \hat{\mathcal{H}}$ is self-adjoint \cite{lemos} with respect
to the internal product,

\begin{equation}
(\Psi ,\Phi ) = \int_0^{\infty} da\, \,\Psi(a,\tau)^*\, \Phi (a,\tau)\, ,
\label{7}
\end{equation}
if the wave functions are restricted to the set of those satisfying either 
$\Psi (0,\tau )=0$ or $\Psi^{\prime}(0, \tau)=0$, where the prime $\prime$
means the partial derivative with respect to $a$. Here, we consider wave 
functions satisfying the former type of boundary condition and we also 
demand that they vanish when $a$ goes to $\infty$.

In order to introduce the noncommutativity in the present model, we shall
follow the prescription used in Refs. \cite{garcia,nelson,barbosa,gil2}. In the
present model, the noncommutativity will be between the two operators $\hat{a}$ 
and $\hat{\tau}$,
\begin{equation}
\label{8}
\left[ \breve{a}, \breve{\tau} \right] = i\theta\, ,
\end{equation}
where $\breve{a}$ and $\breve{\tau}$ are the noncommutative version of the
operators. This noncommutativity between those operators can be taken
to functions that depend on the noncommutative version of those operators with the 
aid of the Moyal product \cite{moyal,bayen,witten,douglas}. Consider 
two functions of $\breve{a}$ and $\breve{\tau}$, 
let's say, $f$ and $g$. Then, the Moyal product between those two function is given by: 
$f(\breve{a},\breve{\tau}) \star g(\breve{a},\breve{\tau}) = f(\breve{a},\breve{\tau})
\exp{\left[(i\theta/2)(\overleftarrow{\partial_{\breve{a}}}\overrightarrow{\partial_{\breve{\tau}}} -
\overleftarrow{\partial_{\breve{\tau}}}\overrightarrow{\partial_{\breve{a}}})\right]}g(\breve{a},\breve{\tau})$.

Using the Moyal product, we may adopt the following Wheeler-DeWitt equation
for the noncommutative version of the present model,
\begin{equation}
\label{9}
\left[\frac{\breve{p}_{\breve{a}}^2}{12} + \breve{p}_{\breve{\tau}}\right] \star \Psi(\breve{a},\breve{\tau}) +
3\breve{a}^2 \star \Psi(\breve{a},\breve{\tau}) = 0.
\end{equation}

It is possible to rewrite the Wheeler-DeWitt equation (\ref{9}) in terms of the commutative version
of the operators $\breve{a}$ and $\breve{\tau}$ and the ordinary product of functions. In order to do that,
we must initially introduce the following transformation between the noncommutative and the commutative 
operators,

\begin{eqnarray}
\label{10}
\breve{a} & = & \hat{a} - \frac{\theta}{2} \hat{p}_{\hat{\tau}},\\
\breve{\tau} & = & \hat{\tau} - \frac{\theta}{2} \hat{p}_{\hat{a}},\nonumber
\end{eqnarray}
and the momenta remain the same. Then, using the properties of the Moyal product it is possible to write 
the potential term in Eq. (\ref{9}) in the following way,
\begin{equation}
\label{11}
3\breve{a}^2 \star \Psi(\breve{a},\breve{\tau}) = 3\left(\hat{a} - 
\frac{\theta}{2} \hat{p}_{\hat{\tau}}\right)^2 \Psi(\hat{a},\hat{\tau}).
\end{equation}
Finally, we may write the commutative version of the Wheeler-DeWitt equation (\ref{9}), to first order
in the commutative parameter $\theta$, in the Schr\"{o}dinger picture as,

\begin{equation}
\label{12}
\frac{1}{12} \frac{\partial^2 \Psi(a,\tau)}{\partial a^2} - 3 a^2 \Psi(a,\tau) =
 - i ( 1 - 3\theta a)\frac{\partial \Psi(a,\tau)}{\partial \tau}.
\end{equation}
For a vanishing $\theta$ this equation reduces to the Schr\"{o}dinger equation of an one dimensional harmonic 
oscillator restricted to the positive domain of the variable \cite{lemos}.

In order to solve this equation, we start imposing that the wave function $\Psi(a,\tau)$ has the 
following form,
\begin{equation}
\label{13}
\Psi(a,\tau) = A(a)e^{-iE\tau}.
\end{equation}
Introducing this ansatz in  Eq. (\ref{12}), we obtain the eigenvalue equation,

\begin{equation}
\label{14}
\frac{d^2 A}{da^2} - 36 a^2 A + (12 - 36\theta a)EA = 0,
\end{equation}
where $E$ is the eigenvalue and it is associated with the fluid energy. It is possible to rewrite 
Eq. (\ref{14}) such that it becomes similar to a one dimensional, quantum mechanical, harmonic 
oscillator eigenvalue equation. In order to do that one has to perform the following transformations,
\begin{eqnarray}
\label{15}
x & = & \sqrt{6} a + 3\theta E/ \sqrt{6},\\\nonumber
\lambda & = & 3\theta^2 E^2/2 + 2E.
\end{eqnarray}
Introducing these transformations in Eq. (\ref{14}), we obtain the new eigenvalue equation,
\begin{equation}
\label{16}
\frac{d^2 A}{dx^2} + (\lambda - x^2)A = 0.
\end{equation}
This equation is the one dimensional, quantum mechanical, harmonic oscillator eigenvalue equation
and has solutions for the following discrete values of $\lambda$,
\begin{equation}
\label{17}
\lambda = 2n + 1,
\end{equation}
where $n = 0, 1, 2, 3,...$. As a consequence of the second transformation of Eq. (\ref{15}) combined
with the result from Eq. (\ref{17}), we obtain that the fluid energy is discrete. It has the
following values,
\begin{equation}
\label{18}
E(n,\theta) = \frac{2}{3\theta^2}\left( -1 + \sqrt{1 + \frac{3}{2}(2n + 1)\theta^2}\right).
\end{equation}
It is important to notice that $E(n,\theta)$ is always positive for any value of $n$ and $\theta$, including 
negative values of $\theta$. Another important property of $E(n,\theta)$ is that, when $\theta \to 0$, in Eq. (\ref{18}), 
we have that $E(n,\theta) \to n + 1/2$. Which is the correct expression for the one dimensional, quantum mechanical, 
harmonic oscillator energies. For a fixed value of $\theta$, $E(n,\theta)$ increases when $n$ increases. On the other
hand, for a fixed value of $n$, $E(n,\theta)$ decreases when $\theta$ increases.

The eigenfunctions to the eigenvalue equation (\ref{16}) are given by,
\begin{equation}
\label{19}
A_n(x) = C_n H_n(x) e^{-x^2/2},
\end{equation}
where $H(x)$ are the Hermite polynomials of degree $n$ and $C_n$ are 
constants. Since we want to consider solutions that vanish at the origin, 
we shall take only the odd degree Hermite polynomials: $n = 1, 3, 5, ...$. 
Besides that, those solutions will be normalized, in the variable 
domain $(0, \infty)$, only if the constants $C_n$ are equal to: 
$C_n = 2^{(n-1)/2} (n!)^{-1/2} (\Pi)^{-1/4}$.

We may, now, write a set of normalized solutions to equation (\ref{12}),
with the aid of the first transformation of Eq. (\ref{15}) and Eq. (\ref{19}),
\begin{equation}
\label{20}
\Psi(a,\tau) = C_n H_n(\sqrt{6} a + 3\theta E(n,\theta)/ \sqrt{6}) e^{-(\sqrt{6} a + 
3\theta E(n,\theta)/ \sqrt{6})^2/2} e^{-iE(n,\theta)\tau}.
\end{equation}

The most general expression of $\Psi(a,\tau)$ Eq. (\ref{13}), which is a solution to Eq. (\ref{12}), is a 
linear combination of the solutions Eq. (\ref{20}), for different values of $n$ for a fixed $\theta$.
It is not difficult to see that, no linear combination of the solutions Eq. (\ref{20}) can satisfy the
boundary condition $\Psi (0,\tau )=0$ (or even $\Psi^{\prime}(0, \tau)=0$), for all $\tau$ except if $E=0$
or $\theta=0$. If $\theta=0$ we return to the commutative case and the case $E=0$ has already been treated
in \cite{hawking}, \cite{nivaldo}. Therefore, we conclude that it is not possible to obtain a wavefunction
satisfying the appropriate boundary conditions for the present model with the considered noncommutativity.

{\bf Acknowledgements.} 
G. A. Monerat thank UERJ for the Prociencia grant.

\end{document}